\documentclass[aps,prl,twocolumn,superscriptaddress]{revtex4}
\usepackage[]{graphicx}

\begin{document}

\title{Influence of Noise on Force Measurements}

\author{Giovanni Volpe}
\email{g.volpe@physik.uni-stuttgart.de}
\affiliation{Max-Planck-Institut f\"{u}r Metallforschung, Heisenbergstra{\ss}e 3, 70569 Stuttgart, Germany}
\affiliation{2. Physikalisches Institut, Universit\"{a}t Stuttgart, Pfaffenwaldring 57, 70550 Stuttgart, Germany}

\author{Laurent Helden}
\affiliation{2. Physikalisches Institut, Universit\"{a}t Stuttgart, Pfaffenwaldring 57, 70550 Stuttgart, Germany}

\author{Thomas Brettschneider}
\affiliation{2. Physikalisches Institut, Universit\"{a}t Stuttgart, Pfaffenwaldring 57, 70550 Stuttgart, Germany}

\author{Jan Wehr}
\affiliation{Department of Mathematics, University of Arizona, Tucson, AZ 85721-0089, USA}

\author{Clemens Bechinger}
\affiliation{Max-Planck-Institut f\"{u}r Metallforschung, Heisenbergstra{\ss}e 3, 70569 Stuttgart, Germany}
\affiliation{2. Physikalisches Institut, Universit\"{a}t Stuttgart, Pfaffenwaldring 57, 70550 Stuttgart, Germany}

\date{\today}

\begin{abstract}
We demonstrate how the ineluctable presence of thermal noise alters the measurement of forces acting on microscopic and nanoscopic objects.
We quantify this effect exemplarily for a Brownian particle near a wall subjected to gravitational and electrostatic forces.
Our results demonstrate that the force measurement process is prone to artifacts if the noise is not correctly taken into account.
\end{abstract}


\maketitle

The concept of force plays a central role in our understanding of nature. Due to the ongoing trend towards miniaturization, the investigation of forces relevant at microscopic and nanoscopic length scales is attracting an increasing amount of attention.
Examples range from the elastic properties of biomolecules \cite{Molecules} to Casimir forces \cite{Casimir}.
Instrumental in such trend has been the invention of new methods to measure ultra-small forces in a range down to few femtonewtons \cite{AFM,PFM}.
Apart from the technological challenge intrinsic to measuring such minute forces,
it is important to realize that the general concept of how forces are measured in macroscopic systems cannot be simply scaled down to
nanoscopic objects, mainly due to the presence of thermal noise affecting the motion of small objects (Brownian motion).
As will be demonstrated below, force measurements in the presence of thermal noise are prone to artifacts, unless the noise is properly taken into account.
However, despite the great number of experiments measuring forces in microscopic systems, the role of noise has not been adequately addressed yet.

In this Letter, we measure the forces acting on a Brownian particle in front of a wall: using two widely employed force-measurement methods, we obtain strongly contrasting forces, which deviate in their magnitude and even their sign.
We track this disagreement down to a well-known mathematical controversy about the interpretation of stochastic differential equations in the presence of a diffusion gradient and we demonstrate how experimental data should be analyzed in order to obtain the correct underlying forces and to avoid artefacts.

\begin{figure}
\includegraphics[width=7.8cm]{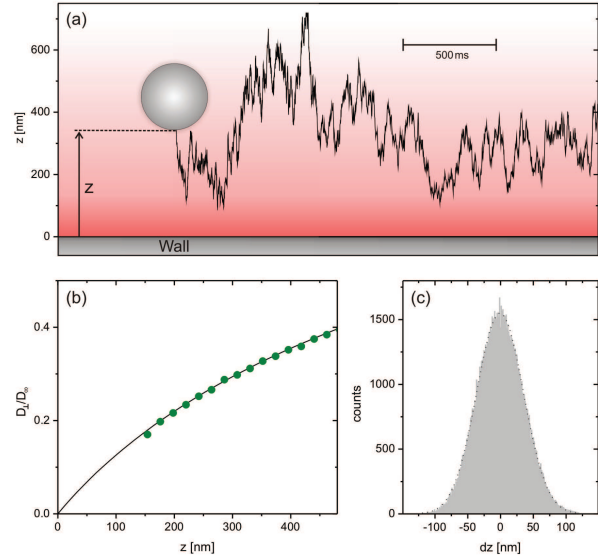}
\caption{(a) A Brownian particle (drawn not to scale) diffuses
near a wall in the presence of gravitional and
electrostatic forces. Its trajectory perpendicular to the wall
is measured with TIRM.
(b) Comparison of measured (bullets) and calculated (line) vertical diffusion coefficient as a function of the particle-wall distance.
(c) Experimentally determined probability distribution of the local drift $dz$ for $dt = 5\, \mathrm{ms}$ at $z = 380\, nm$ (grey).
The dashed line is a gaussian in excellent agreement with the experimental data.\label{fig1}}
\end{figure}

\begin{figure}
\includegraphics[width=7.8cm]{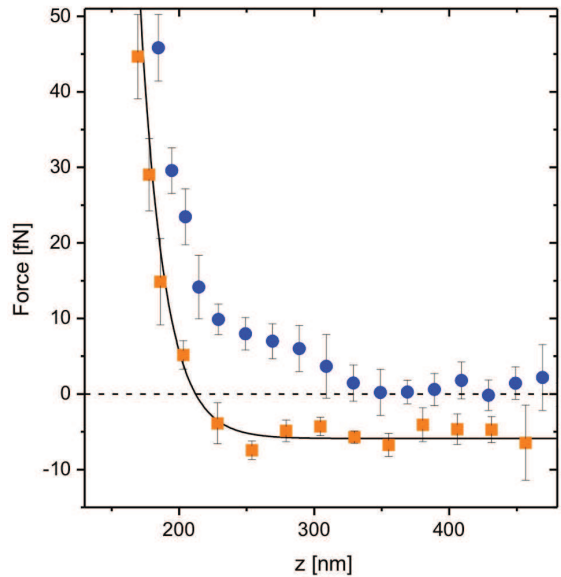}
\caption{Comparison of forces obtained from local drift velocity measurements of a Brownian particle according to
Eq. (\ref{eq:F}) (bullets) and measured from the equilibrium particle height distribution (squares). The solid line corresponds to a fit to Eq. (\ref{Eq:fit}).\label{fig2}}
\end{figure}

When a microscopic body is suspended in a liquid, viscous
forces prevail by several orders of magnitude over inertial
effects \cite{HappelBrenner}. The
presence of a constant external force $F$ in such an overdamped system results in a constant particle drift velocity $v_d = F/\gamma$, where $\gamma$ is the particle's friction coefficient.
Since $v_d = dz/dt$ can be retrieved from the measured particle displacement $dz$ within time $dt$, the force can accordingly be measured as $F = \gamma v_d$.
In the case of large forces this obviously leads to a univocal result.  
However, when the force exerted on the object is comparable or even smaller than the random forces due to the permanent collisions with the surrounding liquid molecules, a different value of the force is measured in each experiment.
Typically one can average over many independent measurements
\begin{equation}\label{eq:F}
\bar{F} = \gamma \bar{v}_d \mbox{,}
\end{equation}
where we define $\bar{v}_d=\frac{1}{dt}\langle dz_j \rangle$ and $j$ denotes the $j$-th experimental value.

Indeed Eq. (\ref{eq:F}) is key in non-equilibrium force measurements, 
e.g. studying the relaxation of biomolecules \cite{Quake2000,PhysRevE.71.061920,PhysicaA_2005_345_173},
colloidal interactions \cite{PhysRevLett.79.4862,PhysRevLett.103.138301}, or
Brownian motion in non-conservative force fields \cite{0295-5075-86-3-38002,PhysRevLett.103.108101,Valentin}.
It must be emphasized, however, that Eq. (\ref{eq:F}) is only valid for a spatially homogeneous diffusion coefficient $D$ of the object to which the force is applied; within linear response this is related to its friction by $D=k_BT/\gamma$ \cite{Valentin}. Generally, Eq. (\ref{eq:F}) does not hold when $D$ varies with position, e.g. due to hydrodynamic interactions between the particle and nearby walls or other particles, a situation often encountered in experiments. It has been shown that such spatial variations in the diffusion coefficient have to be explicitly considered when numerically computing the particle trajectory \cite{Ermak,Clark}. As a consequence, application of Eq. (\ref{eq:F}) leads to erroneous forces, which may severely affect the physical interpretations of experimental data.

To demonstrate the effect of noise on force measurements, we experimentally study a colloidal particle (radius $R=1.31 \pm 0.04 \, \mathrm{\mu m}$, density $\rho_p = 1.51\, \mathrm{g/cm^3}$, MF-F-1.3, Microparticles GmbH) immersed in water (density $\rho_s=1.00\, \mathrm{g/cm^3}$) and diffusing in a closed sample cell above a planar wall, placed at $z=0$ (Fig. \ref{fig1}(a)). This is arguably the simplest realization of a diffusion gradient.

The particle's trajectory perpendicular to the wall $z(t)$ (Fig.  \ref{fig1}(a)) is sampled with nanometer resolution at a sampling rate of $1\, \mathrm{kHz}$ over 200 minutes employing a single particle evanescent light scattering technique known as total internal reflection microscopy (TIRM) \cite{Prieve1999,opticsExpress}. A p-polarized laser beam
($\lambda=658\, \mathrm{nm}$) is totally internally reflected at a
glass-liquid interface generating an evanescent field decaying
into the liquid. 
The particles trajectory is obtained from the scattering intensities, 
which depends on its position relative to the interface.

All conservative forces acting on the particle are \cite{Prieve1999}
\begin{equation}\label{Eq:fit}
F(z) = \,B e^{-\kappa z}-G_{eff}. \label{dlvo}
\end{equation}
The first term is due to
double layer particle-wall forces, with $\kappa^{-1}=18\, \mathrm{nm}$ the Debye length ($300\, \mathrm{\mu M \; NaCl}$ salt) and $B$ a
prefactor depending on the surface charge densities.
The second term accounts for the effective gravitational contribution
$G_{eff} = \frac{4}{3}\pi R^3 (\rho_p - \rho_s) g$, with $g$ the
gravitational acceleration constant.

Far away from any surface, the diffusion coefficient of a spherical particle is $D_{\infty} = k_BT/ 6 \pi\eta R$, where $\eta$ is the  liquid shear viscosity.
Close to a wall, however, the diffusion coefficient sharply decreases due to hydrodynamic interactions. From the solution of the corresponding creeping flow equations one obtains an analytical expression for $D_{\bot}(z)$, the component of $D$ perpendicular to the wall \cite{Brenner1961}, which is plotted for our experimental conditions as solid line in Fig. \ref{fig1}(b). The corresponding data obtained from the experimentally measured particle trajectory according to the conditional average $D_{\bot}(z) = \frac{1}{2dt}\left\langle [z(t+dt) - z(t)]^2 \mid z(t)=z \right\rangle$ \cite{Oetama2005,Carbajal-Tinoco2007} (symbols in Fig. \ref{fig1}c) show excellent agreement with the theoretical prediction; due to the particle-wall electrostatic repulsion only distances above $180\, \mathrm{nm}$ are sampled. 

Since in our system the force depends on $z$, the average drift velocity $ \bar{v}_d$ in Eq. (\ref{eq:F}) has to be replaced by its local value
$\bar{v}_d(z) = {\frac{1}{dt}}\left\langle z(t+dt)-z(t) \mid z(t) = z \right\rangle$. The time interval $dt$ for which the displacement is considered has to be sufficiently small to guarantee that the force acting on the particle can be treated as locally constant. In our experiments this condition is met for $dt \leq 10 \mathrm{ms}$.
Fig. \ref{fig1}(c) shows the probability distribution of  $v_d$ for $dt = 5\, \mathrm{ms}$ and $z = 380\, \mathrm{nm}$, which almost perfectly agrees with a gaussian distribution and thus confirms that within such small time steps, the spatial variation of the force can be neglected. After having replaced the constant friction coefficient in Eq. (\ref{eq:F}) with its local value $\gamma (z)=k_BT/D_{\bot}(z)$ one finally obtains the local force $\bar{F}(z)$ acting on the particle. The result is shown as bullets in Fig. \ref{fig2}.

Since our system is in thermal equilibrium, the forces acting on the particle can be also obtained from the measured particle-wall interaction potential $U(z) = -k_B T \ln p(z)$, where $p(z)$ is the experimental particle's position equilibrium distribution. Contrary to $\bar{F}(z)$, this approach is valid independently of the additional presence of hydrodynamic interactions.
The corresponding force $F(z)=-\frac{d}{dz}U(z)$ is shown as squares in Fig. \ref{fig2} and systematic deviations from $\bar{F}(z)$ are evident. 
In addition, the force-distance relation obtained via the $p(z)$ is in quantitative agreement with Eq. (\ref{Eq:fit}) (solid line in Fig. \ref{fig2}),
where $G_{eff}$ and $\kappa^{-1}$ are taken from the experimentally known parameters. The prefactor $B$ has been treated as a fit parameter and its value $B=770\, \mathrm{pN}$ is in good agreement with other TIRM experiments under similar conditions \cite{Prieve1999}.
In the following, we discuss the reason for this discrepancy providing a simple method to correctly interpret the results of non-equilibrium measurements and to reliably measure forces.

The motion of a Brownian particle can be described by a stochastic differential equation (SDE) where a random function is added to an ordinary differential
equation (ODE) \cite{KS}. This approach was introduced at the beginning of the $\mathrm{20^{th}}$ century
by Smoluchowski, Einstein, Langevin, and Kolmogorov \cite{Ein05,Lan08,Kolmogorov} and put on a firmer mathematical ground in
the 1950s and 1960s by It\={o} and StratonovichÊ\cite{Ito,stra66}. For a Brownian particle
in the presence of a
variable diffusion
coefficient $D_{\bot}(z)$, e.g. near to a wall,
the corresponding SDE reads
\begin{equation}\label{Eq:SDE}
dz = \frac{F(z)}{\gamma(z)}dt + \sqrt{2 D_{\bot}(z)}dW,
\end{equation}
where $W$ is a Wiener process, i.e. a stochastic process almost
surely continuous, almost nowhere differentiable, and whose
increments $dW$ are stationary, independent and normally distributed \cite{KS}.
Integration of Eq. (\ref{Eq:SDE}) yields
\begin{equation}\label{Eq:Integral}
z(T) = z(0) + \int_{0}^{T} \frac{F(z)}{\gamma(z)}
dt+\int_{0}^{T} \sqrt{2 D_{\bot}(z)} dW.
\end{equation}
Due to the irregularity of the Wiener process the value of the stochastic integral on the r.h.s. is ambiguous. It is
defined as the limit of integral sums where the integrand is evaluated inside each bin at a given position parametrized by $\alpha \in [0,1]$,
i.e. $\left. \int_{0}^{T} \sqrt{2 D_{\bot}(z)}dW \right|_{\alpha}= \lim_{N \to \infty}
\sum_{n=0}^{N} \sqrt{2 D_{\bot}(z(t_n))}\Delta W_n$, where $t_n =
\frac{n+\alpha}{N} T$. Since $W$ is a
function of unbounded variation, differently from ordinary
Riemann-Stieltjes integrals, this leads to different
values for each choice of $\alpha$.
A loose understanding of such indetermination can be gained by considering $W$ as 
a random sequence of pulses, each having an infinitesimal duration, but still a finite amplitude; the value of $\alpha$ determines at which position during each jump the integrand should be evaluated \cite{vanKampen}.
Common choices are
$\alpha=0$ (the {\it It\={o} integral}), $\alpha=0.5$ ({\it
Stratonovitch integral}), and $\alpha=1$ ({\it anti-It\={o}} or
{\it isothermal integral}). This is in sharp contrast to ODEs, which have a univocal interpretation.

The values of stochastic integrals for different $\alpha$
are related to each other by a precise mathematical relationship \cite{SDE}.  For example, the solution of Eq. (\ref{Eq:Integral}) can always be written as an It\={o} integral to which an $\alpha$-dependent correction term is added
\begin{equation}\label{Eq:ModifiedIto}
z(T) = z(0) + \int_{0}^{T} \frac{F(z)}{\gamma(z)}dt + 
\left. \int_{0}^{T} \sqrt{2
D_{\bot}(z)}dW \right|_{0}
+ \alpha\int_{0}^{T} \frac{d D_{\bot}(z)}{dz} dt.
\end{equation}
From this one obtains the particle drift velocity
\begin{equation}\label{Eq:Drift}
\bar{v}_d(z)\equiv \frac{\bar{F}(z)}{\gamma(z)}=\frac{F(z)}{\gamma(z)} + \alpha
\frac{d D_{\bot}(z)}{dz}.
\end{equation}
The first term on the r.h.s.  is the deterministic drift due to the ``real" forces
acting on the particle, while the second term represents a  {\it noise-induced drift}. Obviously, the latter only disappears when the diffusion coefficient is homogeneous; otherwise, it has to be accounted for in deducing the forces $F(z)$ acting on a particle from its measured drift velocity $\bar{v}_d(z)$. We remark that, while an equilibrium measurement of the potential constitutes a unambiguous means to experimentally determine the forces, in non-equilibrium situations it is not {\it a priori} clear which value of $\alpha$ should be used and therefore it is not {\it a priori} clear how such noise-induced drift has to be considered in non-equilibrium experiments.

\begin{figure}
\includegraphics[width=7.8cm]{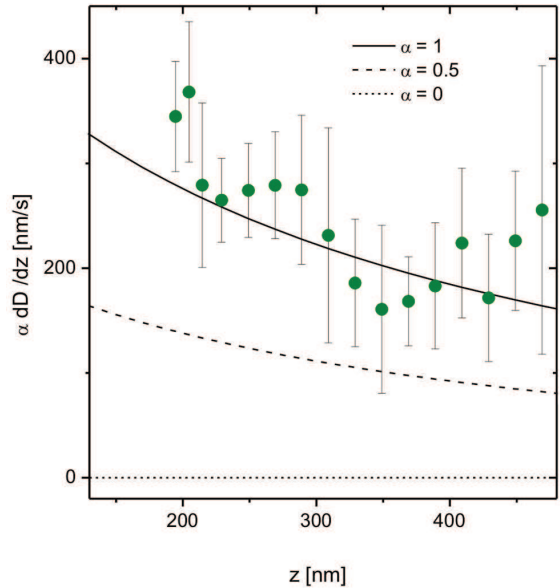}
\caption{Calculated noise-induced drift (lines) for different values of $\alpha$. The symbols correspond to the experimentally determined noise-induced drift.\label{fig3}}
\end{figure}

\begin{figure}
\includegraphics[width=7.8cm]{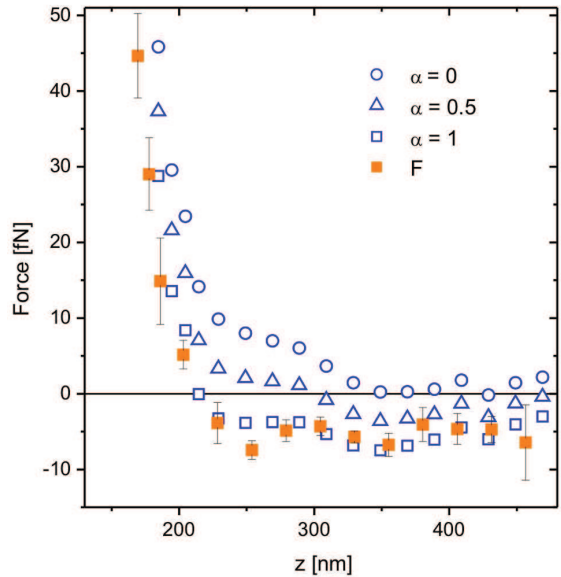}
\caption{Forces obtained from a drift velocity experiment with added noise-induced drift (see Eq. (\ref{eq:Fcorrect}) with $\alpha = 1$ (open squares), $\alpha = 0.5$ (open triangles), and $\alpha = 0$ (open dots). The solid squares represent the forces obtained from an equilibrium measurement (same as in Fig. \ref{fig2}).\label{fig4}}
\end{figure}

From the theoretical $D_{\bot}(z)$ \cite{HappelBrenner} one immediately obtains the noise-induced drift $\alpha \frac{d D_{\bot}(z)}{dz}$, which is plotted as lines for $\alpha=0, 0.5, 1$ in Fig. \ref{fig3}. In order to determine which value of $\alpha$ is valid in our experiment, we have calculated the noise-induced drift from the experimentally measured forces. According to Eq. (\ref{Eq:Drift}) it is determined from the difference of the forces obtained from the drift-velocity measurement and from the equilibrium potential measurement, both presented in Fig. \ref{fig2}, i.e. $(\bar{F}(z)-F(z))/\gamma (z)$. The experimental data (symbols in Fig. \ref{fig3}) show good agreement with the noise-induced drift obtained for $\alpha =1$, i.e. the isothermal integral. All other choices of $\alpha$, in particular negligence of noise-induced drift $(\alpha =0)$, lead to significant differences.

In Fig. \ref{fig4} we have plotted as open symbols
\begin{equation}\label{eq:Fcorrect}
\bar{F}(z) - \alpha \gamma(z) \frac{d D_{\bot}(z)}{dz}\mbox{,}
\end{equation}
i.e. the experimental $\bar{F}(z)$ (open symbols in Fig. \ref{fig2}) with a correction term for
$\alpha=1$ (squares), $\alpha=0.5$ (triangles), and $\alpha=0$ (bullets), respectively. It should be emphasized that not only the absolute value but even the sign of the force depends on the choice of $\alpha$. As closed symbols we have superimposed the forces $F(z)$ obtained from the particle probability distribution (same as solid symbols in Fig. \ref{fig2}), which shows good agreement for $\alpha = 1$. Since the gradient of the diffusion coefficient vanishes far away from the surface, the force-dependence on $\alpha$ is most pronounced close to the wall but weakens at larger $z$.

Our experiments clearly demonstrate that the presence of noise-induced drift has to be considered in non-equilibrium force measurements; otherwise, this can lead to artifacts in the measured forces, which may even suggest the wrong sign of the force. While the correction is in our case on the order of several femtonewton, it becomes more significant in the presence of larger diffusion gradients, i.e. for shorter particle-wall distances or for smaller particles.
We stress, furthermore, that a constant diffusion can be assumed only for a particle far from any boundary.
Such boundaries are naturally introduced by surfaces or by other particles in suspension, a situation that is typically met in experiments.

In the cases in which thermodynamic consistency must be satisfied, $\alpha=1$ is the correct choice \cite{Lancon2001,Lau2007}.
This is particularly true for the experimental system we have investigated, i.e. a colloidal particle performing Brownian motion coupled to a thermal bath.
Indeed, the convention $\alpha = 1$ is the only one {\it naturally} leading
to the usually accepted steady-state probability distribution; 
other conventions require the addition of a {\it spurious drift} term to the Langevin equation to account for the noise-induced drift \cite{Lau2007}.
One is also {\it naturally} led to the convention $\alpha = 1$ when
considering the vanishing mass limit of the second-order
Smolukowski equation, which, despite containing a random term, has
an unambiguous interpretation even in the presence of a diffusion
dependent on position, but not on velocity. We remark, however, that in case
of a velocity-dependent diffusion coefficient also the
interpretation of the Smolukowski equation becomes ambiguous
\cite{klimontovich}.

In a more general sense the value of $\alpha$ depends on the system under
study \cite{Kupferman2004}.
What works for
the motion of a Brownian particle might not be appropriate for the
description of other stochastic processes, e.g. stock market behavior or 
ecosystem dynamics.
This study does not claim to find out which is the ``right" value
of $\alpha$  for all the situations modeled by SDEs.
Nonetheless, it demonstrates that the intrinsic
ambiguity of SDEs with multiplicative noise is indeed amenable to
experimental scrutiny and it may clear the way for similar studies
in other fields that make an intensive use of SDEs, such as economics
and biology.

\begin{acknowledgments}
We thank Thomas Franosch, Udo Seifert, Peter Reimann, Andrea
Gambassi, Gerhard N\"agele, Markus Rauscher, Valentin Blickle, and Jakob Mehl for helpful and
stimulating discussions.
\end{acknowledgments}


\end{document}